\documentclass[twocolumn,10pt]{tsfp}
\usepackage{flushend}
\usepackage{graphicx}
\usepackage{soul} 
\usepackage{booktabs}
\usepackage{color}
\usepackage[authoryear,round]{natbib}
\usepackage{fancyhdr}
\usepackage{lineno}
\usepackage{color}
\usepackage{graphicx}
\usepackage{booktabs}
\usepackage{booktabs}
\usepackage{subfig}
\usepackage{soul} 
\usepackage{layouts}
\usepackage{lscape}
\usepackage{mathtools} 
\usepackage{relsize}   
\usepackage{xcolor}
\usepackage{tikz}
\usepackage{floatrow}
\usepackage{natbib}
\usepackage{amsmath}
\usepackage{commath}
\usepackage{enumitem}
\usepackage{newtxtext}
\usepackage{newtxmath}
\usepackage{natbib}
	
\newcommand\Rey{\mbox{\textit{Re}}}  
\newcommand\Fro{\mbox{\textit{Fr}}}  
\newcommand{\appropto}{\mathrel{\vcenter{
			\offinterlineskip\halign{\hfil$##$\cr
				\propto\cr\noalign{\kern2pt}\sim\cr\noalign{\kern-2pt}}}}}

\graphicspath{ {./figures/}}

\vspace{-10mm}

\title{Flow past an inclined spheroid in homogeneous and stratified environments}

\author{Sheel Nidhan
	\affiliation{Mechanical and Aerospace Engineering\\
		University of California San Diego\\
		CA 92093\\
		snidhan@ucsd.edu
	}	
}

\author{Jose L. Ortiz-Tarin
	\affiliation{Mechanical and Aerospace Engineering\\
		University of California San Diego\\
		CA 92093\\
		jlortiztarin@gmail.com
	}
}

\vspace{-5mm}
\author{Sutanu Sarkar
	\affiliation{Mechanical and Aerospace Engineering\\
		University of California San Diego\\
		CA 92093\\
		sarkar@ucsd.edu
	}
}

\begin{document}
	
	\maketitle   
	\thispagestyle{fancy}
	
	\fontsize{9}{11}\selectfont
	
	\section*{ABSTRACT}
	
	Large eddy simulations (LES) are performed to study the flow past a 6:1 prolate spheroid placed at an angle of incidence of $\alpha = 10^{\circ}$. The diameter-based Reynolds number ($\Rey = U_{\infty}D/\nu$) is set to a value of $5\times10^{3}$ and four values of diameter-based Froude numbers ($\Fro = U_\infty/ND$) are analyzed: $\Fro = \infty, 6, 1.9,$ and $1$. Visualizations of coefficient of pressure ($C_p$) and friction ($C_f$) contours reveal asymmetry in the $\Fro= \infty$ and $6$ flows while, at $\Fro = 1$ and $\Fro = 1.9$, the flow over body does not have any visible asymmetry. This finding is further corroborated through the analysis of force coefficients on the body. The changes in the pressure coefficients ($C_p$), friction coefficients ($C_f$), and drag coefficients ($C_d$) with the Froude number are described in detail for $\alpha = 10^{\circ}$. We also present the analyses of forces on the body at $\alpha = 0^{\circ}$ angle of incidence for comparison with the $\alpha = 10^{\circ}$ cases.
	
	\vspace{-5mm}
	\section*{INTRODUCTION}
	
	Despite their widespread presence in  applications, slender body flows have received significantly less attention compared to the flows past bluff bodies. There are some particularities that make the study of slender body flows especially challenging. In a numerical simulation, resolving the boundary layer (BL) of a long body is computationally more expensive than resolving the BL of a blunt body. Additionally, in experimental studies, slender bodies can take up a significant portion of the measurement section and, since they normally generate thin wakes, they are potentially harder to probe and measure. 
	
	The first experimental study of flow past a slender body dates back to \cite{Chevray1968} who investigated the wake of a 6:1 prolate spheroid at $\Rey = 4.5 \times 10^{5}$. The wake measurements spanned a streamwise distance of $x/D = 18$. \cite{han1979flow} conducted an experimental study of flow past a 4.3:1 spheroid at different angles of incidence ($\alpha$) at $\Rey \approx 2 \times 10^{4}$. They primarily focused on the flow separation pattern, identifying two regimes as $\alpha$ was changed: (i) closed or bubble separation and (ii) open or free-vortex type separation. \cite{wang1990three} extended the study of \cite{han1979flow} to prolate spheroids with aspect ratios of 2:1, 3:1, and 4:1 for a wide range of $\alpha$  in the range $[0^{\circ}, 90^{\circ}]$. 
	Experimental studies have also been performed at higher $\Rey$ with tripped BL by \cite{fu1994flow} and \cite{chesnakas1994full}. \cite{Jimenez2010} and \cite{ashok_asymmetries_2015} studied the high $\Rey$ wake of a DARPA SUBOFF at $\alpha = 0^{\circ}$ and in pitch configurations, respectively. While \cite{Jimenez2010} focused on the self-similarity and scalings in the near to intermediate wake, \cite{ashok_asymmetries_2015} characterized the wake asymmetries induced due to the pitch configuration.
	
	Numerical simulations have also been used to study slender body flows. \cite{constantinescu2002numerical} and \cite{wikstrom2004large} conducted RANS and LES studies of flow past a 6:1 prolate spheroid at $\Rey = 4.2 \times 10^6$ and $\alpha = 10^{\circ}$, $20^{\circ}$. \cite{tezuka_three-dimensional_2006} carried out a three-dimensional stability analysis study for the flow past a 4:1 spheroid for varying $\alpha$. For $\alpha \neq 0^{\circ}$, as $\Rey$ was increased, they found transition from a symmetric to an asymmetric flow configuration in the longitudinal center-plane. The flow asymmetry at nonzero angle of incidence was later studied in more detail using DNS by \cite{jiang_transitional_2015}. In the last few years, the wake of a DARPA SUBOFF has also been studied through LES by \cite{Posa2016,Kumar2018} at $\Rey \sim O(10^{5})$. \cite{ortiz2021high} was the first study of an unstratified slender body flow that probed  far wake statistics, extending to $x/D = 80$.
	
	The LES study by \cite{Ortiz-tarin_stratified_2019} is the first to take background stratification into account for the study of slender body flows. Their high-resolution LES study was conducted for a 4:1 spheroid at zero angle of attack, $\Rey = 10^{4}$ and $\Fro = \infty, 3, 1$ and $0.5$. They analyzed the laminar BL evolution, force distribution, and the near- and far-field characteristics of the steady lee waves. In the current work, we build upon their work studying the effect of varying $\Fro$ on the flow past a 6:1 spheroid placed at a moderate angle of incidence ($\alpha = 10^{\circ}$). We analyze the effect of stratification on: (i) the variation of $C_p$, $C_f$, (ii) the forces on the body, and (iii) the flow separation. We also present a brief analysis of forces on the body at $\alpha=0^{\circ}$ for comparison with $\alpha = 10^{\circ}$ cases. To the best of our knowledge, this is the first study exploring the flow characteristics of an inclined slender body flow in stratified environments. In future work, we plan to extend this study to the analysis of vorticity and wake dynamics in these flows.
	
	\vspace{-5mm}
	\section*{NUMERICAL METHODOLOGY}
	
	The numerical solver used for these simulations have been extensively validated in the past for body-inclusive simulations of stratified wakes \citep{Ortiz-tarin_stratified_2019,chongsiripinyo_decay_2020,ortiz2021high,nidhan_dynamic_2019,nidhan2020spectral,nidhan2022analysis}. For a detailed description of the solver, boundary conditions and immersed boundary method (IBM) approach, we refer the interested readers to \cite{chongsiripinyo_decay_2020}.
	
	In the present work, $\Rey = 5 \times 10^{3}$ and  four different $\Fro = \infty, 6,1.9,$ and $1$ are simulated for two angles of incidence $\alpha = 0^{\circ}$ and $10^{\circ}$. For the stratified cases at $\alpha = 10^{\circ}$, radial and streamwise domains span  $0 \leq r/D \leq 53$ and $-30 \leq x/D \leq 50$, respectively. A large radial extent together with a sponge layer on the boundaries weaken the IGWs before they hit the end of the computational domain and hence control the amplitude of spurious reflected waves. Numbers of grid points in different directions are as follows: $N_{r} = 1000$ in the  radial direction, $N_{\theta} = 128$ in the azimuthal direction, and $N_{x} = 3584$ in the streamwise direction. For the unstratified wake at $\alpha = 10^{\circ}$, radial and streamwise domains span $0 \leq r/D \leq 21$ and $-11 \leq x/D \leq 47$ while $N_r = 718, N_{\theta} = 256$, and $N_x = 2560$. At $\alpha = 0^{\circ}$, we simulate the wakes till a downstream distance of $x/D =30$ and up to $r/D = 17$ and $53$ for unstratified and stratified cases, respectively. The grid point distribution for $\alpha = 0^{\circ}$ cases are as follows: $N_r = 1000$ and $910$ for stratified and unstratified cases, respectively, $N_\theta$ = 128, and $N_x = 3584$. 
	\vspace{-5mm}
	\section*{BODY FORCES AT $\alpha = 0^{\circ}$}
	
	\begin{figure*}[ht]
		\centering
		\includegraphics[width=5.5in,keepaspectratio]{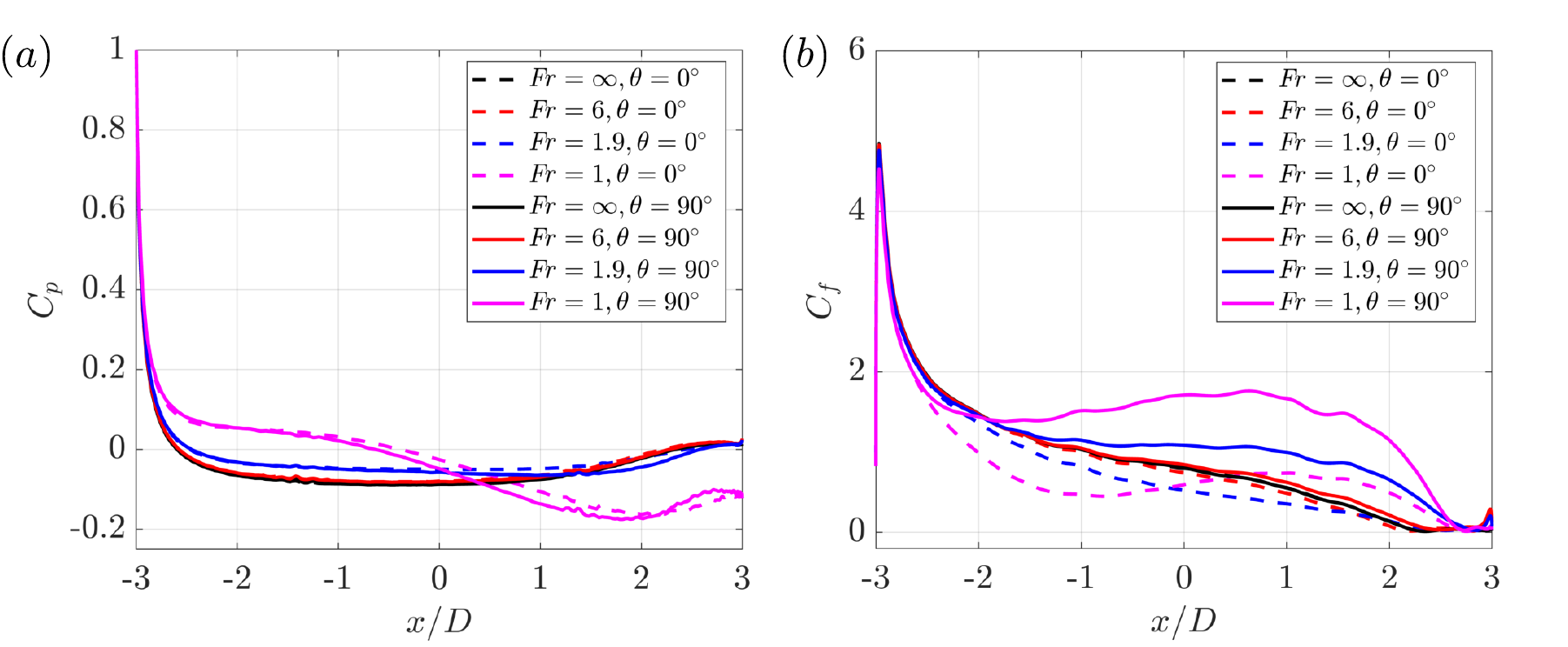}
		\caption{Variation of (a) pressure coefficient $C_p$ and (b) skin-friction coefficient $C_f$ for different $\Fro$ at zero angle of attack ($\alpha = 0^{\circ}$). The $\theta = 0^{\circ}$ and $90^{\circ}$ curves correspond to  variations on the surface in the horizontal ($\theta = 0^{\circ}$) plane and the vertical ($\theta=90^{\circ}$) planes on the spheroid. Plotted value of $C_f$ is $\Rey^{0.5}$ times  the friction coefficient.} 
		\label{fig:cp_cf_aoa0}
	\end{figure*}
	
	Figure \ref{fig:cp_cf_aoa0}(a) and (b) show the variation of the pressure coefficient $C_p = (P-P_{\infty})/ 0.5\rho U_\infty^2$ and skin-friction coefficient $C_f = \Rey^{0.5} |\tau_x|/0.5\rho U_{\infty}^2$, respectively, for different values of  $\Fro$ at $\alpha = 0^{\circ}$. In the reported $C_f$,  the raw skin friction coefficient is multiplied by  $ \Rey^{0.5}$  to obtain  an  $ O(1)$ value. The  $\theta = 0^{\circ}$ and $90^{\circ}$ labels correspond to the horizontal $x-y$ and vertical $x-z$ plane, respectively.  An averaging time window of $30D/U_\infty$ is used to obtain these results.
	
	For $\Fro \geq 1.9$ cases (figure \ref{fig:cp_cf_aoa0}(a)), the drop of pressure and its recovery primarily occur towards the beginning and the  end of the body, respectively. For $-2 \lessapprox x/D \lessapprox 1$, $C_p$ remains approximately constant for $\Fro \geq 1.9$. On the other hand, $C_p$ in $\Fro = 1$ case shows a monotonic decay till $x/D \approx 2$ and a slight recovery for $x/D \gtrapprox 2$, indicating strong effect of buoyancy on the flow over the body. For $\Fro  \geq  1.9$, $C_p$ variations are qualitatively very similar. The differences between $\Fro = \infty$ and the critical $\Fro_c$ curves are less pronounced in the 6:1 spheroid ($\Fro_c = 1.9$) than what was observed in the 4:1 spheroid ($\Fro_c \approx 1$) as reported by \cite{Ortiz-tarin_stratified_2019}. Anisotropy between horizontal and vertical plane $C_p$ curves appears at $\Fro = 1.9$ and increases slightly at $\Fro = 1$. Pressure visualizations (not shown here) confirm that the low pressure region in the tail of the spheroid (in the vertical plane) at $\Fro = 1$ is imposed by the steady lee-wave field, indicating a strong influence of buoyancy on the flow at this value of $\Fro$. 
	
	Figure \ref{fig:cp_cf_aoa0}(b) shows the variation of $C_f$ for different $\Fro$ at $\alpha = 0^{\circ}$. Similar to the behavior of $C_p$, the variation of  $C_f$ for $\Fro = 6$ and $\infty$ are  very similar. The flow separates at $x/D = 2.34$ for the unstratified flow (marked by $C_f \rightarrow 0$). This value is in excellent agreement with the result of \cite{patel_topology_1994}. $\Fro = 1.9$ shows elevated and suppressed levels of $C_f$ in vertical and horizontal planes, respectively, compared to $\Fro = 6$ and $\infty$. At $\Fro = 1$, $C_f$ in the vertical plane further increases compared to $\Fro = 1.9$. Moreover, in the horizontal plane as well, $\Fro = 1$ shows higher $C_f$ than $\Fro = \infty$ and $6$ for $x/D \gtrapprox 0.5$. Increased $C_f$ in the vertical plane for $\Fro = 1$ and $1.9$ is a consequence of thinner BL over the body (not shown here for brevity) compared to $\Fro = 6$ and $\infty$. In the horizontal plane, for $\Fro = 1$ (compared to $\Fro =\infty$), BL thickens between $x/D \approx -2$ to $0$ and gets thinner beyond $x/D \approx 1$, explaining the trend of $C_{f}$ in the horizontal plane for $\Fro = 1$.
	
	\begin{table}
		\centering
		\caption{Drag coefficients ($C_d$) and corresponding pressure ($C_d^{p}$) and friction contributions ($C_d^{f}$) for $\alpha = 0^{\circ}$ at different $\Fro$.}
		\label{tab:cd_aoa0}
		\begin{tabular}{c c c c c}
			& & \\ 
			\hline
			\hline
			$\alpha=0^{\circ}$, $\Fro$ & $C_d$	& $C_d^{f}$ & $C_d^{p}$ & $\Delta C_d = C_d - C_d (\infty)$ \\
			\hline
			$\Fro = \infty$ &  0.24  & 0.22  & 0.02 & 0    \\
			$\Fro = 6$      &  0.26  & 0.23  & 0.03 & 0.02 \\
			$\Fro = 1.9$    &  0.32  & 0.25  & 0.07 & 0.08 \\
			$\Fro = 1$      &  0.52  & 0.29  & 0.23 & 0.28 \\
			\hline
			\hline
		\end{tabular}
	\end{table}
	
	Table \ref{tab:cd_aoa0} presents $C_d = F_{d}/(0.5\rho_oU_{\infty}^2A)$ for different $\Fro$ at $\alpha = 0^{\circ}$. Here, $A = \pi D^2/4$. There is a monotonic increase in $C_d$, $C_d^{p}$, and $C_d^{f}$ with increasing stratification levels. Friction contributes more to the drag than pressure (except at $\Fro = 1$), as expected for a slender body flow. However, compared to flow past a 4:1 spheroid \citep{Ortiz-tarin_stratified_2019}, we find that the effect of stratification on the overall drag, quantified by $\Delta C_d = C_d - C_d(\infty)$, is weaker in the 6:1 spheroid. Between $\Fro = \infty$ and $\Fro =1.9$, $C_d$ changes by $\approx 33\%$ in the present case while there was a $100\%$ increase in the 4:1 spheroid. This smaller increase in $C_d$ for 6:1 spheroid is primarily due to a smaller increase in $C_d^{p}$ at $\Fro_c = 1.9$ (compared to $\Fro = \infty$), unlike in the 4:1 prolate spheroid. It is only when $\Fro = 1$ that we see a sharp jump in $C_d^{p}$ leading to a $\approx 100\%$ increase in $C_d$, consistent with the differences we observe in $C_p$ curves (figure \ref{fig:cp_cf_aoa0}a) between $\Fro = 1$ and $\Fro \geq 1.9$ cases. 
	
	It is a well-established observation that the amplitude of steady lee-waves decreases with increasing value of
	$\Fro$ in  the $\Fro \geq 1$ regime
	for a variety of wake generators, including 6:1 prolate spheroid \citep{bonneton_internal_1993,meunier2018internal}. Specifically, \cite{meunier2018internal} found that the lee-wave amplitude for a 6:1 spheroid (based on $\partial w/\partial z$) decayed as $\Fro^{-2}$. On the other hand, it should be noted that the definition of critical $\Fro_c$ given by \cite{Ortiz-tarin_stratified_2019} is based purely on kinematic considerations, i.e., by equating half-wavelength of the lee-wave to the length of the body. As a result, $\Fro_c = L/\pi D$ increases linearly with the aspect ratio. Thus,  for the 6:1 spheroid, we see a weaker effect of stratification on the drag at its critical value of $\Fro_c = 6/\pi = 1.9 $ compared to the 4:1 spheroid for which $\Fro_c = 4/\pi =  1.27 $ is lower.  That the  lee wave field in the $\Fro = 1.9$ case is weaker  than in the $\Fro = 1$ case is also confirmed also by quantification of the pressure field, not shown here for brevity. 
	
	\vspace{-5mm}
	\section*{BODY FORCES AT $\alpha = 10^{\circ}$}
	
	Introducing even a moderate angle of incidence, namely $\alpha = 10^{\circ}$, significantly changes the characteristics of flow on the body, reflected by significantly different trends of $C_p$, $C_f$ and force coefficients in $\alpha =10^{\circ}$ flows compared to their $\alpha = 0^{\circ}$ counterparts. In what follows, we discuss the trends of above-mentioned quantities when $\alpha = 10^{\circ}$ and $\Fro = \infty, 6, 1.9$, and $1$. For $\alpha = 10^{\circ}$, a time averaging window of approximately $50D/U_\infty$ is used.
	
	\vspace{-5mm}
	\subsection*{Coefficient of Pressure $C_{p}$}
	
	\begin{figure*}[ht]
		\centering
		\includegraphics[width=5.75in,keepaspectratio]{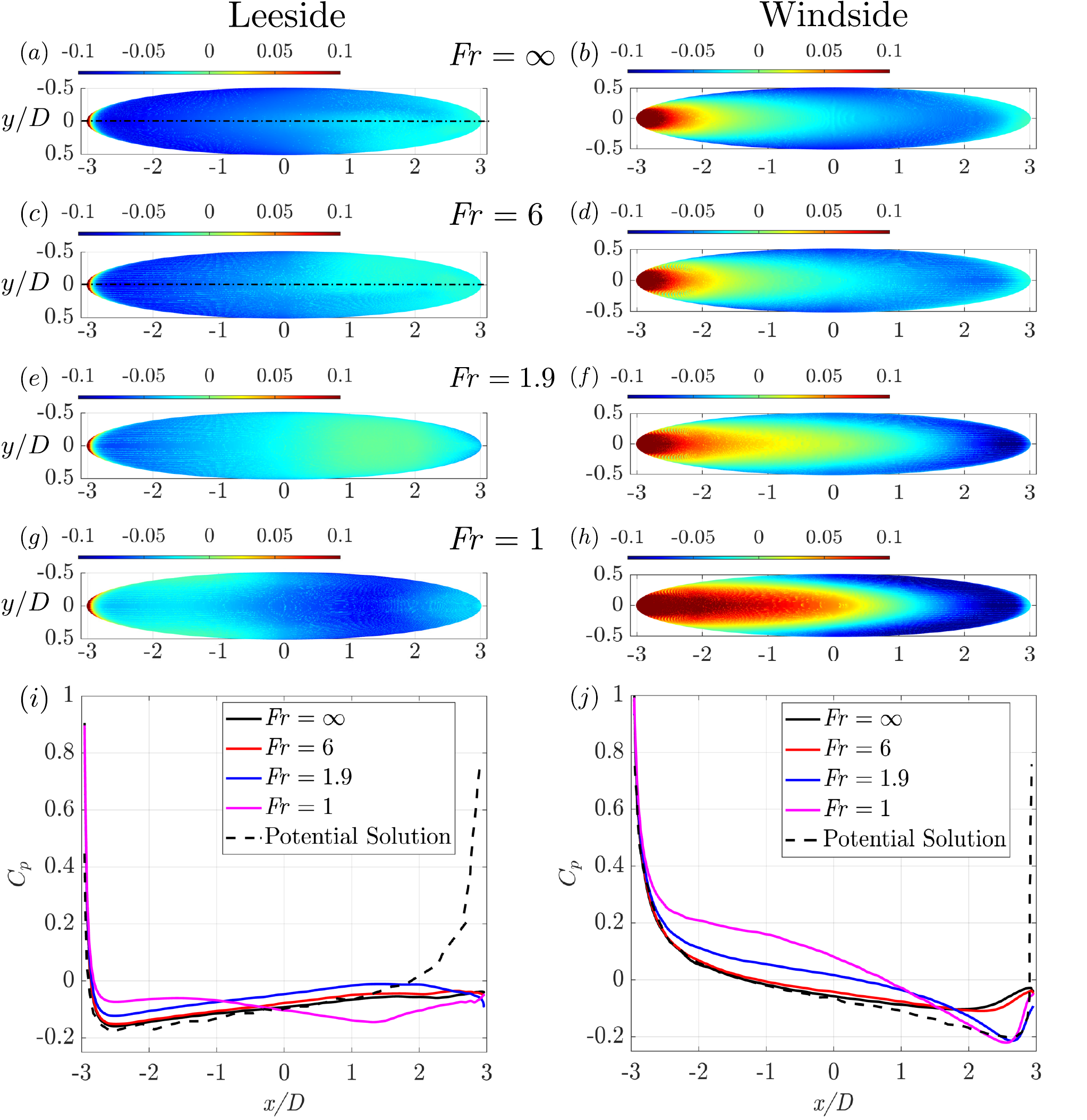}
		\caption{Pressure contours on the leeside (a,c,e)  and windside (b,d,f)  of the spheroid shown for all simulated $\Fro$ at $\alpha = 10^{\circ}$. Dashed lines in (a) and (c) correspond to $y=0$.  Also shown is the variation of $C_p$ (i, j) on the leeside and windside of the body surface in the $y=0$ plane. Potential solution for $C_p$ is also shown in dashed line, obtained from \cite{piquet_navier-stokes_nodate}.} 
		\label{fig:cp_contours_aoa10}
	\end{figure*}
	
	\begin{figure*}[ht]
		\centering
		\includegraphics[width=5.75in,keepaspectratio]{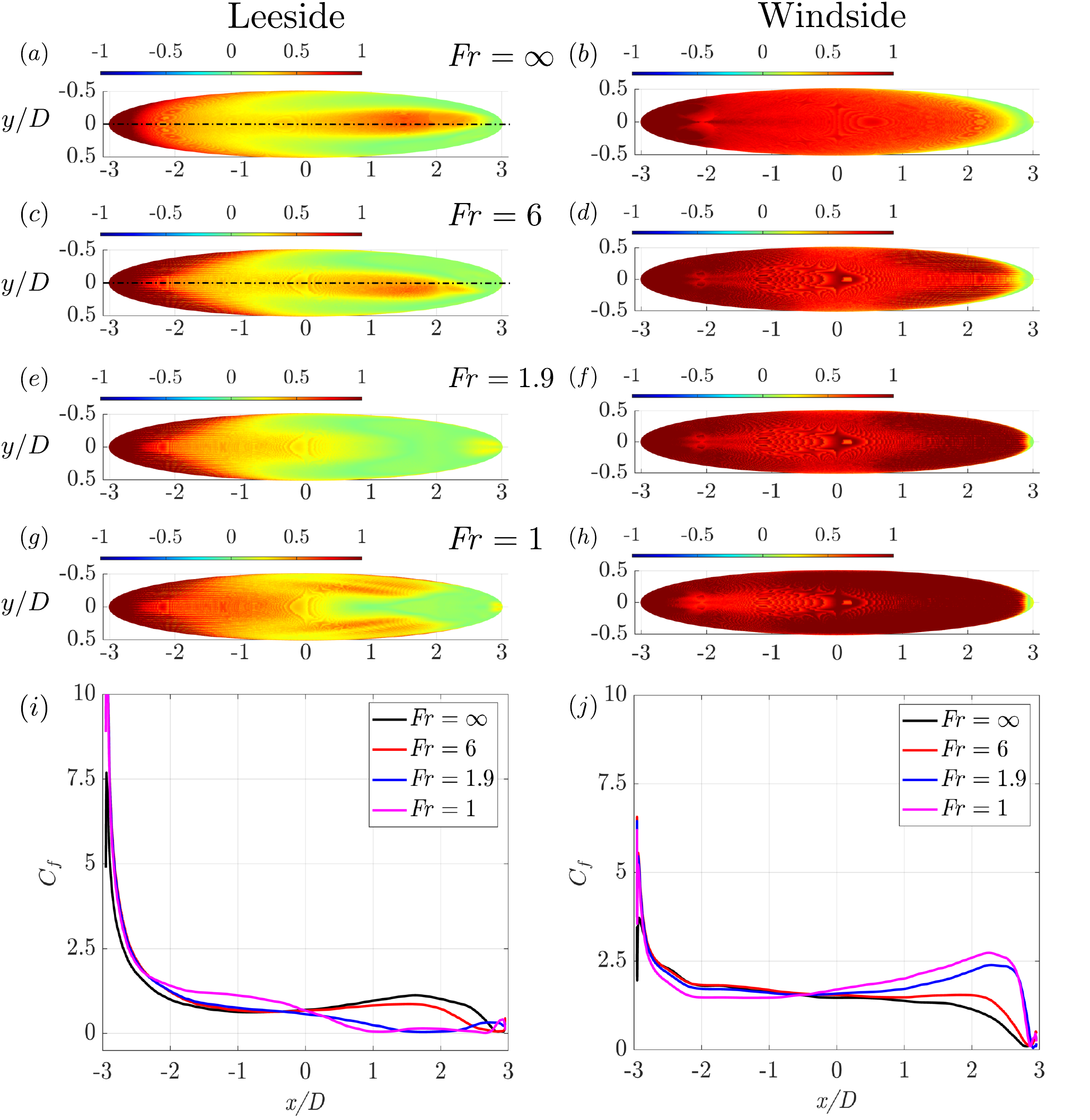}
		\caption{Contours of $\Rey^{0.5} |\tau_x|$ (a-h) on the leeside and windside of the spheroid for all $\Fro$ at $\alpha = 10^{\circ}$. Dashed lines in (a) and (c) correspond to $y=0$. Variation of $C_f$ (i, j) on the leeside and windside of the body at $y=0$ plane.} 
		\label{fig:cf_contours_aoa10}
	\end{figure*}
	
	\begin{figure*}[ht]
		\centering
		\includegraphics[width=5.5in,keepaspectratio]{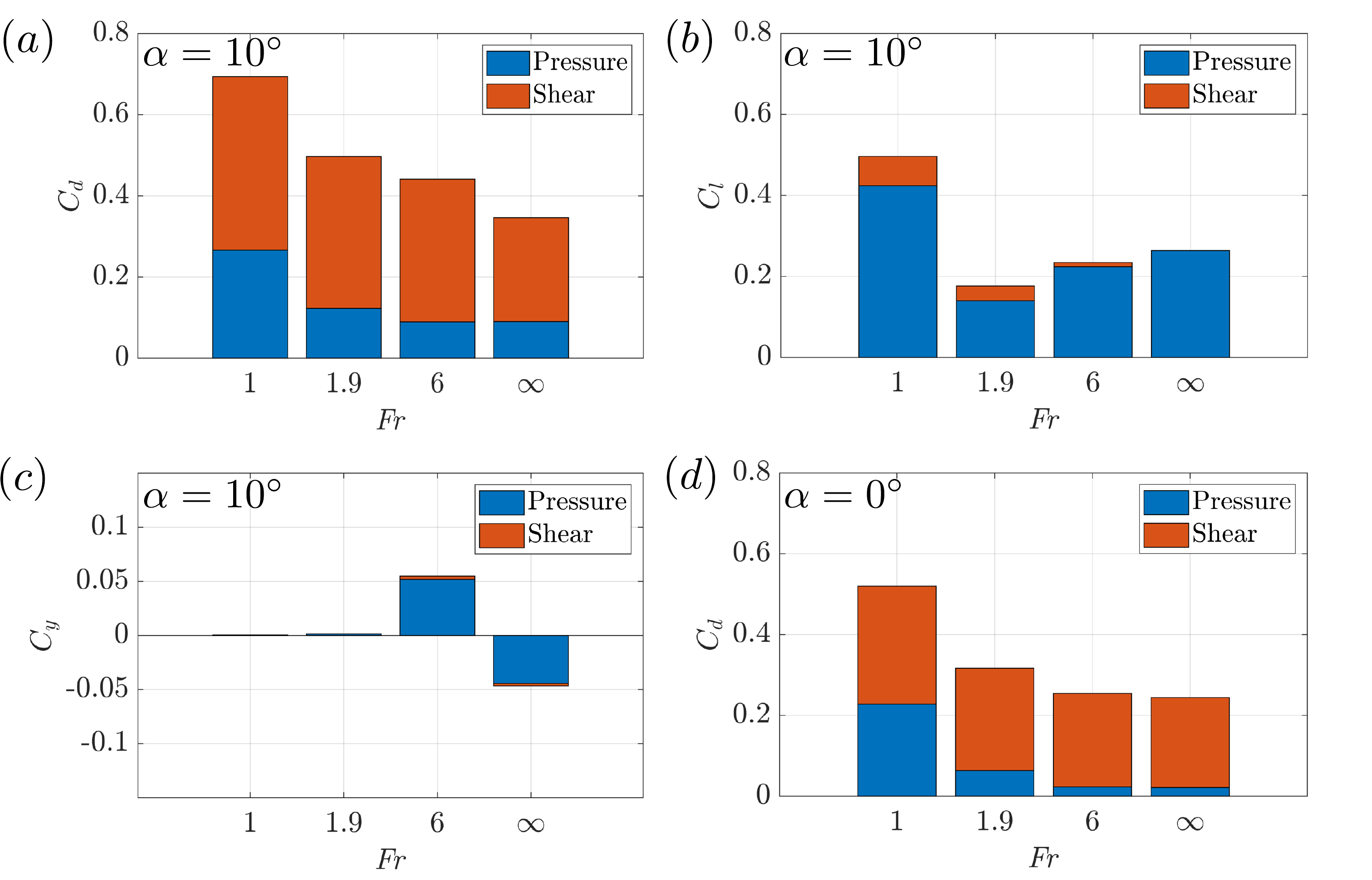}
		\caption{Force coefficients decomposed between pressure and shear contribution: (a) coefficient of drag $C_d$, (b) coefficient of lift $C_l$, (c) lateral force $C_y$ at $\alpha =10^{\circ}$, and (d) $C_d$ at $\alpha = 0^{\circ}$.} 
		\label{fig:force_partition_plots}
	\end{figure*}
	
	Figure \ref{fig:cp_contours_aoa10}(a-h) show the pressure contours on the leeside and windside of the body. We also present the variation of $C_p$ along $y=0$ line on the leeside and windside in figure \ref{fig:cp_contours_aoa10}(i,j), respectively. $C_p$ obtained from potential flow solution is also presented for comparison in figure \ref{fig:cp_contours_aoa10}(i,j). On both sides, the agreement between the potential solutions and LES simulations is excellent till $x/D \approx 1$. Beyond $x/D \approx 1$ LES simulations deviate from potential solutions, presumably due to the three-dimensionality of BL evolution and effects of flow separation. Pressure drop on both sides primarily happens near to the nose of the body and in the ascending order of $\Fro$, i.e., the pressure drop in the nose for $\Fro = 1 > 1.9 > 6 > \infty$ (figure \ref{fig:cp_contours_aoa10}(i)). It is also worth noting that the $\Fro = 6$ and $\infty$ cases show asymmetry in the leeside pressure contours about the $y=0$ plane as shown in figure \ref{fig:cp_contours_aoa10}(a,c).
	
	On the leeside, for $\Fro \geq 1.9$ cases, pressure recovers continuously, albeit slowly, after the initial drop in the nose (figure \ref{fig:cp_contours_aoa10}i). On the contrary, pressure in the $\Fro = 1$ case, which shows the lowest initial drop (from the stagnation value at the nose)   of all cases in the fore of  the body ($x/D < 0$) drops below that of the remaining cases for $x/D > 0$. This is evident both from contours (figure \ref{fig:cp_contours_aoa10}g) and $C_p$ plots (figure \ref{fig:cp_contours_aoa10}i). Mean pressure contours in the vicinity of the spheroid (not shown here for brevity), reveal a strong low-pressure region on the entirety of the leeside. This low-pressure region is imposed by the steady lee wave field at  $\Fro = 1$. The differences between the unstratified case and strongly stratified cases ($\Fro = 1.9$ and $1$) are even more pronounced on the windside compared to the leeside of the spheroid. Beyond $x/D \approx 2$, $C_p$ of $\Fro = 1.9$ and $1$ fall significantly below those of $\Fro = \infty$ and $6$ (figure \ref{fig:cp_contours_aoa10}j), followed by a sharp recovery towards the end. The contours in figure \ref{fig:cp_contours_aoa10}(b,d,f,h) confirm this trend of higher and lower pressure at the head and tail, respectively, for the $\Fro = 1$ and $1.9$ cases compared to $\Fro \geq 6$. 
	
	\vspace{-5mm}
	\subsection*{Coefficient of Friction $C_{f}$}
	
	Figure \ref{fig:cf_contours_aoa10}(a-h) show the contours of scaled shear stress $\Rey^{0.5} |\tau_x|$ on the leeside (left) and windside (right) for all cases at $\alpha = 10^{\circ}$. In figure \ref{fig:cf_contours_aoa10}(i,j), we present the variation of $C_f$ on the spheroid surface and  in the $y=0$ plane on both sides. Similar to the pressure contours, a distinct asymmetry is present on the leeside of $\Fro = \infty$ and $6$ flows as shown figure \ref{fig:cf_contours_aoa10}(a,c), to be discussed in more detail in the next subsection. 
	
	The shear stress contours show that flow separation primarily happens on the leeside at $\alpha = 10^{\circ}$ across all $\Fro$. The region of separated flow can be identified by the region of $|\tau_x| \rightarrow 0$, marked in green in figure \ref{fig:cf_contours_aoa10} contours. For $\Fro = \infty$ and $6$, flow separates primarily from the two lateral sides on the lee of the body (figure \ref{fig:cf_contours_aoa10}(a,c)). There is a central region near $y =0$ which remains attached until nearly the tail. In the $\Fro = 1.9$ case, separation occurs from the sides as well as  the  central region around the $y=0$ plane (figure \ref{fig:cf_contours_aoa10}e) while in the $\Fro = 1$ case, separation primarily happens in the central region (figure \ref{fig:cf_contours_aoa10}g) and not at the sides. Thus, it can be inferred that the stratification level strongly influences the flow separation even at the moderate non-zero angles of attack of this study.
	
	In the vertical-center plane ($y=0$),  $C_f$ varies similarly for all $\Fro$ on both sides (figure \ref{fig:cf_contours_aoa10}(i,j)) till $x/D \approx 0$. On the leeside (figure \ref{fig:cf_contours_aoa10}i) and for $x/D \geq 0$, the $\Fro = \infty$ and $6$ cases show higher $C_f$ than the strongly stratified cases of $\Fro = 1.9$ and $1$. On the windside, beyond $x/D \approx 1$, $\Fro \leq 1.9$ cases show elevated $C_f$ as compared to $\Fro \geq 6$ (figure \ref{fig:cf_contours_aoa10}j). This region of elevated surface shear in the $\Fro = 1.9$ and $1$ flows coincides with the region of steep pressure drop observed at $x/D > 1$ (figure \ref{fig:cp_contours_aoa10}j). 
	
	\vspace{-5mm}
	\subsection*{Force Coefficients}
	
	Figure \ref{fig:force_partition_plots}(a,b,c) present the force coefficients ($C_i = F_i/0.5\rho U_\infty^2 A$) at $\alpha = 10^{\circ}$, decomposed between pressure and friction contributions. Here $C_x, C_y$ and $C_z$ correspond to drag ($C_d$), lateral force, and lift ($C_l$) on the body, respectively. We first discuss $C_d$ and $C_l$ and follow by noting the unusual characteristics of $C_y$ for $\alpha = 10^{\circ}$. Figure \ref{fig:force_partition_plots}(d) also shows $C_d$ for $\alpha = 0^{\circ}$ for reference. In $\alpha = 0^{\circ}$ cases, $C_l \approx 0$ and  $C_y \approx 0$ at all $\Fro$ indicating no asymmetry in the flow over body. 
	
	For all $\Fro$, $C_d$, $C_d^{f}$ and $C_d^{p}$ (figure \ref{fig:force_partition_plots}a) at $\alpha = 10^{\circ}$ increase relative to the zero degree angle of incidence ((figure \ref{fig:force_partition_plots}d). Similar to $\alpha = 0^{\circ}$, there is a weak monotonic increase in $C_d$ till $\Fro = 1.9$ and a large jump thereafter at $\Fro = 1$. This jump primarily comes from an approximately $100\%$ increase in $C_d^{p}$ from $\Fro = 1.9$ to $\Fro = 1$. Figure \ref{fig:cp_contours_aoa10}(g,h) show that the pressure contours at $\Fro = 1$ are quite different from those of $\Fro \geq 1.9$. The tail and nose on the leeside and windside, respectively, are at a lower and higher pressure (compared to $\Fro \geq 1.9$ cases) which leads to an enhanced $C_d^{p}$. For $C_l$, primary contributor is the pressure rather than shear, as expected for a moderate angle of incidence. It is interesting to note that $C_l$ decreases till $\Fro = 1.9$ and then increases significantly at $\Fro = 1$ case, resulting from an increase in $C_l^{p}$. The reason for this increase is the large difference between the leeside and windside pressure in $\Fro = 1$ flow as shown in figure \ref{fig:cp_contours_aoa10}(g,h).
	
	Figure \ref{fig:force_partition_plots}(c) shows that $C_y \neq 0$ for $\Fro = \infty$ and $6$ at $\alpha = 10^{\circ}$. For $\Fro = \infty$, $C_y \approx -0.05$ and for $\Fro = 6$, $C_y \approx 0.05$. This value is approximately $12\%$ of the streamwise drag force in both cases. Intuitively, $C_y$ should be equal to zero due to reflectional symmetry in the configuration about the $y=0$ plane. Non-zero $C_y$ implies lateral asymmetry in the flow on the body. Pressure and friction contours in figure \ref{fig:cp_contours_aoa10}(a,c) and \ref{fig:cf_contours_aoa10}(a,c), respectively, show that this is indeed the case and the asymmetry originates on the leeside of the body at $x/D \gtrapprox 1$ (the $y =0$ intersection of the body surface is shown to better identify this lateral asymmetry). No asymmetry is present on the windside flow over the spheroid. When the value of $\Fro$ is decreased to $1.9$, $C_y \rightarrow 0$. Further decrease to $\Fro = 1$ also results in $C_y = 0$. This indicates that the flow asymmetry is suppressed as the stratification is increased. Figure \ref{fig:cp_contours_aoa10}(e,g) and figure \ref{fig:cf_contours_aoa10}(e,g) also confirm that for $\Fro = 1.9$ and $1$, no asymmetry is visually evident on the leeside of the spheroid. Hence, two important findings from the analysis of $C_y$ are: (i) the weakly stratified ($\Fro = 6$) and unstratified ($\Fro = \infty$) flow exhibit lateral asymmetry and (ii) this asymmetry is suppressed as the strength of stratification is increased, i.e., at $\Fro = 1.9$ and $1$.
	
	Our finding regarding lateral asymmetry in the unstratified case  is in accord with previous studies on flow past slender bodies \citep{tezuka_three-dimensional_2006,ashok_asymmetries_2015,ashok_structure_2015,jiang_transitional_2015}. It is also interesting to note that the $C_y$ of $\Fro = \infty$ and $\Fro = 6$ are similar in magnitude but flipped in sign. We hypothesize that the $\Fro = \infty$ and $\Fro = 6$ flows might be locked in two different reflectional-symmetry-breaking states, with each state being equally probable. There can be a switching between these two states at a very large timescale. The existence of a long time scale for switching between different reflection-symmetry-breaking states have been extensively researched in flow past three-dimensional blunt bodies \citep{grandemange_turbulent_2013, rigas_low-dimensional_2014, dalla_longa_simulations_2019}. Interestingly, \cite{jiang_transitional_2015}, who also found lateral asymmetry in their flow (6:1 spheroid at $\alpha = 45^{\circ}$), did not find a switch  even after $600D/U_\infty$. We aim to investigate  the characteristics of the intermediate to far wake, besides looking at near-body flow. Hence, running the simulations for $T \sim 1000D/U_{\infty}$ would be prohibitively expensive and out of scope of the current work.
	
	\vspace{-5mm}
	\section*{CONCLUSIONS}
	
	Large eddy simulations (LES) are performed to study the characteristics of flow over a 6:1 spheroid placed at an angle of incidence $\alpha =  10^{\circ}$. We present results at $\Rey = 5\times 10^{3}$ and $\Fro = \infty, 6, 1.9$,  and $1$. Additionally, flows past a 6:1 spheroid at the same $(\Rey,\Fro)$ combinations but at $\alpha=0^{\circ}$ are simulated to provide a basis for comparison.  We find that the buoyancy effect introduced by stratification  strongly modulates the pressure and friction forces on the body. The critical Froude number~\citep{Ortiz-tarin_stratified_2019} given by $\Fro_c = L/\pi D$ is defined kinematically by equating the body length to the half-wavelength of the generated lee wave.  For the 6:1 spheroid,  the effect of stratification on $C_p,C_f,$ and $C_d$ is more pronounced at $\Fro = 1$ than at the critical Froude number  $\Fro_c \approx 1.9$.  For both $\alpha = 0^{\circ}$ and $10^{\circ}$ cases, $C_d$ monotonically increases with decreasing $\Fro$. At $\alpha = 10^{\circ}$, a distinct lateral asymmetry is visible in the $C_p$ and $C_f$ contours for $\Fro = 6$ and $\infty$ cases. This gives rise to a non-zero lateral force whose magnitude is approximately $12\%$ of the streamwise drag at $\alpha = 10^{\circ}$. Further increasing the stratification  kills the asymmetry at $\Fro = 1.9$ and $1$. At $\alpha = 10^{\circ}$, we also find that the flow separation over the body is strongly dependent on the value of $\Fro$. At $\Fro = 6$ and $\infty$, the flow separates from  the lateral sides on the lee of the body  while at $\Fro =1$, the flow separates predominantly near the vertical center plane $y=0$. In the future studies, we plan to investigate flow separation and  topology of shed vortices and their potential link to the flow asymmetry in $\alpha=10^{\circ}$ cases as well as wake dynamics.
	
	\vspace{-5mm}
	\section*{ACKNOWLEDGMENTS}
	We gratefully acknowledge the support of the Ofﬁce of Naval Research grant N0014-20-1-2253.
	
	\vspace{-5mm}
	\bibliographystyle{tsfp}
	\bibliography{tsfp}

\begin{thebibliography}{27}
\expandafter\ifx\csname natexlab\endcsname\relax\def\natexlab#1{#1}\fi

\bibitem[Ashok {\em et~al.\/}(2015{\natexlab{{\em a\/}}})Ashok, Van~Buren \&
  Smits]{ashok_asymmetries_2015}
Ashok, A., Van~Buren, T. \& Smits, A. J. 2015{\natexlab{{\em a\/}}}
  Asymmetries in the wake of a submarine model in pitch. {\em J. Fluid Mech.\/}
  {\bf 774}, 416--442.

\bibitem[Ashok {\em et~al.\/}(2015{\natexlab{{\em b\/}}})Ashok, Van~Buren \&
  Smits]{ashok_structure_2015}
Ashok, A., Van~Buren, T. \& Smits, A.~J. 2015{\natexlab{{\em b\/}}} The
  structure of the wake generated by a submarine model in yaw. {\em Exp
  Fluids\/} {\bf 56}~(6), 123.

\bibitem[Bonneton {\em et~al.\/}(1993)Bonneton, Chomaz \&
  Hopfinger]{bonneton_internal_1993}
Bonneton, P., Chomaz, J.~M. \& Hopfinger, E.~J. 1993 Internal waves produced by
  the turbulent wake of a sphere moving horizontally in a stratified fluid.
  {\em J. Fluid Mech.\/} {\bf 254}, 23--40.

\bibitem[Chesnakas \& Simpson(1994)]{chesnakas1994full}
Chesnakas, C.J. \& Simpson, R.L. 1994 Full three-dimensional measurements of
  the cross-flow separation region of a 6: 1 prolate spheroid. {\em Exp.
  Fluids\/} {\bf 17}~(1), 68--74.

\bibitem[Chevray(1968)]{Chevray1968}
Chevray, R. 1968 {The turbulent wake of a body of revolution}. {\em J. Basic
  Eng\/} pp. 275--284.

\bibitem[Chongsiripinyo \& Sarkar(2019)]{chongsiripinyo_decay_2020}
Chongsiripinyo, K. \& Sarkar, S. 2019 Decay of turbulent wakes behind a disk in
  homogeneous and stratified fluids. {\em J. Fluid Mech.\/} {\bf 885}, A31.

\bibitem[Constantinescu {\em et~al.\/}(2002)Constantinescu, Pasinato, Wang,
  Forsythe \& Squires]{constantinescu2002numerical}
Constantinescu, G.~S., Pasinato, H., Wang, Y., Forsythe, J.~R. \& Squires,
  K.~D. 2002 Numerical investigation of flow past a prolate spheroid. {\em J.
  Fluids Eng.\/} {\bf 124}~(4), 904--910.

\bibitem[Dalla~Longa {\em et~al.\/}(2019)Dalla~Longa, Evstafyeva \&
  Morgans]{dalla_longa_simulations_2019}
Dalla~Longa, L., Evstafyeva, O. \& Morgans, A.~S. 2019 Simulations of the
  bi-modal wake past three-dimensional blunt bluff bodies. {\em J. Fluid
  Mech.\/} {\bf 866}, 791--809.

\bibitem[Fu {\em et~al.\/}(1994)Fu, Shekarriz, Katz \& Huang]{fu1994flow}
Fu, T.C., Shekarriz, A., Katz, J. \& Huang, T.T. 1994 The flow structure in the
  lee of an inclined 6: 1 prolate spheroid. {\em J. Fluid Mech.\/} {\bf 269},
  79--106.

\bibitem[Grandemange {\em et~al.\/}(2013)Grandemange, Gohlke \&
  Cadot]{grandemange_turbulent_2013}
Grandemange, M., Gohlke, M. \& Cadot, O. 2013 Turbulent wake past a
  three-dimensional blunt body. {Part} 1. {Global} modes and bi-stability. {\em
  J. Fluid Mech.\/} {\bf 722}, 51--84.

\bibitem[Han \& Patel(1979)]{han1979flow}
Han, T. \& Patel, V.~C. 1979 Flow separation on a spheroid at incidence. {\em
  J. Fluid Mech.\/} {\bf 92}~(4), 643--657.

\bibitem[Jiang {\em et~al.\/}(2015)Jiang, Gallardo, Andersson \&
  Zhang]{jiang_transitional_2015}
Jiang, F., Gallardo, J.~P., Andersson, H.~I. \& Zhang, Z. 2015 The transitional
  wake behind an inclined prolate spheroid. {\em Phys. Fluids\/} {\bf 27}~(9),
  093602.

\bibitem[Jim{\'{e}}nez {\em et~al.\/}(2010)Jim{\'{e}}nez, Hultmark \&
  Smits]{Jimenez2010}
Jim{\'{e}}nez, J.~M., Hultmark, M. \& Smits, A.~J. 2010 {The intermediate wake
  of a body of revolution at high Reynolds numbers}. {\em J. Fluid Mech.\/}
  {\bf 659}, 516--539.

\bibitem[Kumar \& Mahesh(2018)]{Kumar2018}
Kumar, P. \& Mahesh, K. 2018 {Large-eddy simulation of flow over an
  axisymmetric body of revolution}. {\em J. Fluid Mech.\/} {\bf 853}, 537--563.

\bibitem[Meunier {\em et~al.\/}(2018)Meunier, Le~Diz{\`e}s, Redekopp \&
  Spedding]{meunier2018internal}
Meunier, P., Le~Diz{\`e}s, S., Redekopp, L. \& Spedding, G.~R. 2018 Internal
  waves generated by a stratified wake: experiment and theory. {\em J. Fluid
  Mech.\/} {\bf 846}, 752--788.

\bibitem[Nidhan {\em et~al.\/}(2020)Nidhan, Chongsiripinyo, Schmidt \&
  Sarkar]{nidhan2020spectral}
Nidhan, S., Chongsiripinyo, K., Schmidt, O.~T. \& Sarkar, S. 2020 Spectral
  proper orthogonal decomposition analysis of the turbulent wake of a disk at
  re= 50 000. {\em Phys. Rev. Fluids\/} {\bf 5}~(12), 124606.

\bibitem[Nidhan {\em et~al.\/}(2019)Nidhan, Ortiz-Tarin, Chongsiripinyo, Sarkar
  \& Schmid]{nidhan_dynamic_2019}
Nidhan, S., Ortiz-Tarin, J.~L., Chongsiripinyo, K., Sarkar, S. \& Schmid, P.~J.
  2019 Dynamic {Mode} {Decomposition} of {Stratified} {Wakes}. In {\em {AIAA}
  {Aviation} 2019 {Forum}\/}. Dallas, Texas: American Institute of Aeronautics
  and Astronautics.

\bibitem[Nidhan {\em et~al.\/}(2022)Nidhan, Schmidt \&
  Sarkar]{nidhan2022analysis}
Nidhan, S., Schmidt, O.T. \& Sarkar, S. 2022 Analysis of coherence in turbulent
  stratified wakes using spectral proper orthogonal decomposition. {\em J.
  Fluid Mech.\/} {\bf 934}.

\bibitem[Ortiz-Tarin {\em et~al.\/}(2021)Ortiz-Tarin, Nidhan \&
  Sarkar]{ortiz2021high}
Ortiz-Tarin, J.L., Nidhan, S. \& Sarkar, S. 2021 High-reynolds-number wake of a
  slender body. {\em J. Fluid Mech.\/} {\bf 918}.

\bibitem[Ortiz-Tarin {\em et~al.\/}(2019)Ortiz-Tarin, Chongsiripinyo \&
  Sarkar]{Ortiz-tarin_stratified_2019}
Ortiz-Tarin, J.~L., Chongsiripinyo, K.~C. \& Sarkar, S. 2019 Stratified flow
  past a prolate spheroid. {\em Phys. Rev. Fluids\/} {\bf 4}~(9), 094803.

\bibitem[Patel \& Kim(1994)]{patel_topology_1994}
Patel, V.C. \& Kim, S.E. 1994 Topology of laminar flow on a spheroid at
  incidence. {\em Comp. Fluids\/} {\bf 23}~(7), 939--953.

\bibitem[Piquet \& Queutey(1992)]{piquet_navier-stokes_nodate}
Piquet, J. \& Queutey, P. 1992 Navier-stokes computations past a prolate
  spheroid at incidence - {I}. {Low} incidence case. {\em Comp. Fluids\/} {\bf
  21}, 599--625.

\bibitem[Posa \& Balaras(2016)]{Posa2016}
Posa, A. \& Balaras, E. 2016 {A numerical investigation of the wake of an
  axisymmetric body with appendages}. {\em J. Fluid Mech.\/} {\bf 792},
  470--498.

\bibitem[Rigas {\em et~al.\/}(2014)Rigas, Oxlade, Morgans \&
  Morrison]{rigas_low-dimensional_2014}
Rigas, G., Oxlade, A.~R., Morgans, A.S. \& Morrison, J.~F. 2014 Low-dimensional
  dynamics of a turbulent axisymmetric wake. {\em J. Fluid Mech.\/} {\bf 755},
  R5.

\bibitem[Tezuka \& Suzuki(2006)]{tezuka_three-dimensional_2006}
Tezuka, Asei \& Suzuki, Kojiro 2006 Three-{Dimensional} {Global} {Linear}
  {Stability} {Analysis} of {Flow} {Around} a {Spheroid}. {\em AIAA Journal\/}
  {\bf 44}~(8), 1697--1708.

\bibitem[Wang {\em et~al.\/}(1990)Wang, Zhou, Hu \& Harrington]{wang1990three}
Wang, K.C., Zhou, H.C., Hu, C.H. \& Harrington, S. 1990 Three-dimensional
  separated flow structure over prolate spheroids. {\em Proc. R. Soc. A: Math.
  Phys. Eng. Sci.\/} {\bf 429}~(1876), 73--90.

\bibitem[Wikstr{\"o}m {\em et~al.\/}(2004)Wikstr{\"o}m, Svennberg, Alin \&
  Fureby]{wikstrom2004large}
Wikstr{\"o}m, N., Svennberg, U., Alin, N. \& Fureby, C. 2004 Large eddy
  simulation of the flow around an inclined prolate spheroid. {\em J.
  Turbul.\/} {\bf 5}~(1), 029.

\end{thebibliography}


\begin{thebibliography}{5}
\expandafter\ifx\csname natexlab\endcsname\relax\def\natexlab#1{#1}\fi

\bibitem[Kwon \& Pletcher(1981)]{Kwon}
Kwon, O.~K. \& Pletcher, R.~H. 1981 Prediction of the incompressible flow over
  a rearward-facing step. Technical Report HTL-26, CFD-4. Iowa State Univ.,
  Ames, IA.

\bibitem[Lee {\em et~al.\/}(1982)Lee, Korpela \& Horne]{Lee}
Lee, Y., Korpela, S.~A. \& Horne, R.~N. 1982 Structure of multi-cellular
  natural convection in a tall vertical annulus. In {\em Proceedings of 7th
  International Heat Transfer Conference\/} (ed. U.~Grigul et~al.), , vol.~2,
  pp. 221--226. Washington, D.C.: Hemisphere Publishing Corp.

\bibitem[Sparrow(1980{\natexlab{{\em a\/}}})]{Sparrow}
Sparrow, E.~M. 1980{\natexlab{{\em a\/}}} Fluid-to-fluid conjugate heat
  transfer for a vertical pipe - internal forced convection and external
  natural convection. {\em ASME Journal of Heat Transfer\/} {\bf 102},
  402--407.

\bibitem[Sparrow(1980{\natexlab{{\em b\/}}})]{Sparrow2}
Sparrow, E.~M. 1980{\natexlab{{\em b\/}}} Forced-convection heat transfer in a
  duct having spanwise-periodic rectangular protuberances. {\em Numerical Heat
  Transfer\/} {\bf 3}, 149--167.

\bibitem[Tung(1982)]{Tung}
Tung, C.~Y. 1982 Evaporative heat transfer in the contact line of a mixture.
  PhD thesis, Rensselaer Polytechnic Institute, Troy, NY.

\end{thebibliography}
	
\end{document}